\documentclass[12pt]{article}
\usepackage{epsfig}
\usepackage{url}
\makeatletter
\def\alt{\mathrel{\mathpalette\gl@align<}}
\def\agt{\mathrel{\mathpalette\gl@align>}}
\def\gl@align#1#2{\lower.6ex\vbox{\baselineskip\z@skip\lineskip\z@
\ialign{$\m@th#1\hfil##\hfil$\crcr#2\crcr\sim\crcr}}}
\makeatother

\begin{document}
\begin{flushright}
%{\tt hep-ph/yymmnnn}\\
MIFPA-13-06\\
UMD-PP-013-002\\
February, 2013
\end{flushright}
\vspace*{1.0cm}
\begin{center}
\baselineskip 20pt
{\Large\bf Proton decay and $\mu\to e+\gamma$ Connection in a Renormalizable SO(10) GUT for Neutrinos
} \vspace{1cm}

{\large
Bhaskar Dutta$^1$, Yukihiro Mimura$^2$ and R. N. Mohapatra$^3$}
\vspace{.5cm}

$^1${\it
Department of Physics, Texas A\&M University,
College Station, TX 77843-4242, USA
}\\
%\vspace{.5cm}
$^2${\it
Department of Physics, National Taiwan University, Taipei,
Taiwan 10617, R.O.C.
}\\
$^3${\it
Department of Physics, University of Maryland, College Park, MD 20742, USA
}\\

\vspace{.5cm}
%$^\dagger${\it
%}\\

\end{center}

\begin{center}{\bf Abstract}\end{center}

Supersymmetric SO(10) grand unified models with renormalizable Yukawa couplings involving {\bf 10}, {\bf 126} and {\bf 120} Higgs fields have been shown to give a very economical theory for understanding quark-lepton flavor in a unified framework. In previous papers, we showed how nucleon decay can be suppressed in these models without invoking cancellation, by choice of Yukawa flavor texture within a type II seesaw framework for neutrinos that explains all mixings and masses  including the recently observed ``large''  $\theta_{13}$. In this follow-up paper, we extend our earlier work to the case of type I seesaw and show  that the recently measured ``large''   $\theta_{13}$ can be accommodated in this case while suppressing proton decay.  We then point out that the two cases (type I and II) lead to different testable predictions for $B(\mu\to e+\gamma)$ and $B(\tau\to \mu (e) +\gamma)$ as well as  different flavor final states in nucleon decay. In particular, we find that for the type I seesaw case, $B(\tau\to \mu +\gamma)$ can be observable while at the same time suppressing $B(\mu\to e+\gamma)$, whereas in the type II seesaw case, $B(\tau\to \mu +\gamma)$ is always suppressed whereas $B(\mu\to e+\gamma)$ is observable.

\thispagestyle{empty}

\bigskip
\newpage

\addtocounter{page}{-1}

\section{Introduction}
\baselineskip 18pt

The last decade has seen incredible experimental progress in the field of neutrino physics. Since the discovery of the oscillations of atmospheric neutrinos in 1998
followed by the confirmation of the solar neutrino oscillations discovered by Ray Davis in mid-sixties, many parameters describing the neutrino masses and mixings
have been measured in various solar, atmospheric and accelerator and reactor experiments. The most recent such measurement is that of the remaining mixing parameter $\theta_{13}$ by reactor as well as accelerator experiments%In fact, the measurements of the remained to be addressed 13 mixing are now reported
~\cite{An:2012eh,Ahn:2012nd,Abe:2011sj,abe,MINOS} which has created a lot of excitement in the theory community due to its implications for physics behind the neutrino masses.
%and 
%the measurement by Daya Bay collaboration (even alone) 
%excludes a zero value with a significance of 7.7 standard deviations \cite{An:2012bu}.
Attention has now turned  to measuring the other missing pieces of informations: Dirac versus Majorana nature of neutrinos,
 the CP violating phase $\delta$ in the lepton sector  as well as  the mass hierarchy of neutrinos. %
The global analysis of the parameter fit \cite{Fogli:2012ua} seem 
to have some preference for CP phase $\sim \pi$ although it is far too early to take this seriously.  
The ``large'' value of the 13 mixing angle (around $\sin\theta_{13} \simeq 0.15$) however, has raised the hope
that the CP phase can be measured in near future.

What have we learnt about the physics behind the neutrino masses from these observations?  The first challenge is to understand why the neutrino masses are so small and the second is to see if the same framework that explains the small neutrino mass also simultaneously explains the observed mixing pattern. A popular paradigm for this seems to be the so-called seesaw mechanism \cite{seesaw} where one extends the standard model by adding heavy Majorana right-handed neutrinos (known as the type I seesaw) or  heavy SM Higgs triplets (known as the type II seesaw)~\cite{Schechter:1980gr}. Generic models based on seesaw are quite successful in achieving the first goal i.e. small neutrino masses. It is however more challenging to get an understanding of the mixings without further assumptions. The two approaches which have been used extensively are: (i) leptonic symmetries and (ii) grand unified theories. The closeness of the leptonic mixing parameters to group theory-like numbers (e.g. $\frac{1}{\sqrt{2}}; \frac{1}{\sqrt{3}}$) have been the major driving force behind the symmetry approach. Most models with symmetries however predicted the third mixing angle $\theta_{13}=0$, which have now been disproved by experiments as noted. While there are still a number of leptonic symmetry scenarios
which can lead to large value of $\theta_{13}$ \cite{feruglio}, the fact that
\begin{equation}
\sin\theta_{13} \sim \sqrt{\frac{\Delta m^2_{12}}{\Delta m^2_{23}}}\sim \theta_{Cabibbo}
\label{13-massratio}
\end{equation}
may be suggestive of a ``big'' picture that unifies quarks and leptons such as those based on grand unified theories. In particular, if normal neutrino hierarchy is established, that would imply a similarity between the quark and the lepton sector, that would be consistent with expectations from grand unified theories (GUTs)
 and would further motivate the GUT approach to all flavors. 

One class of grand unified theories that have been extensively explored and appear quite successful in providing a unified description of quark-lepton flavor i.e. giving observed values of the neutrino mixings and masses, while being consistent with observed quark masses and CKM  mixings
is the one based on the group SO(10) with or without supersymmetry and using {\bf 126} fields to break the $B-L$ subgroup of SO(10). This class of models uses only  renormalizable Yukawa interactions involving a  {\bf 10} and {\bf 126} \cite{Babu:1992ia,Fukuyama:2002ch} with/without an additional {\bf 120}~\cite{Dutta:2004wv,Yang:2004xt} to explain both the quark and lepton flavor puzzle. The renomalizability property restricts the fermion Yukawa couplings to have only a small number of parameters (without the need for adhoc symmetries) describing both the quark and the lepton sectors simultaneously so that the models become quite predictive. Indeed these models led to the prediction not only of the relation between neutrino mass ratios and Cabibbo angle in Eq.(\ref{13-massratio}) but also the value of $\theta_{13}$  several years prior to its measurement \cite{Goh:2003sy} without the need for {\bf 120} field. 

One of the key predictions of the supersymmetric (SUSY) GUT models is the enhancement of proton decay rate due to the presence of new supersymmetric contributions from color triplet Higgsino exchange. These contributions are known to severely constrain the nature of the these theories and clearly if the above SO(10) neutrino models are to be taken seriously, their  consistency with current lower limits on proton lifetimes must be examined for both the decay modes $p\to e^+\pi^0$ where the key input is the value of the unification scale and the characteristic SUSY mode $p\to K^+\bar{\nu}$, where the key input is the nature Higgs fields that give mass to fermions. For the case of minimal SO(10) models with {\bf 10}, {\bf 126}, with type II seesaw for neutrino masses, this consistency requires  cancellation between parameters describing the Higgsino mixings and masses \cite{Goh:2003nv}. Since there are many proton decay modes that have been constrained by experiments \cite{SK} and all parameters of the model except for three Higgsino mixngs are already determined, the fact that there is a consistent picture, is highly nontrivial. However this works only for low tan$\beta$ regime of the MSSM parameter space. It was subsequently pointed out \cite{Dutta:2004zh,Dutta:2007ai} that once the {\bf 120} is included, the situation improved quite a bit within the type II seesaw framework i.e. by choice of an appropriate flavor structure for the different Yukawa coupling matrices, one can not only give a simultaneous fit to fermion masses  but also suppress proton decay to the desired level, for both small and large tan $\beta$, without need to invoke cancellation. It was shown that in the model with {\bf 120}, one can obtain $\theta_{13}$ as large as 0.15, which is in accord with the recent measurements. 

An important question remained as to whether the idea of suppressing proton decay  by choice of flavor structure works in models with {\bf 120} when type I seesaw is picked for understanding neutrino masse. The second question is: is their a way to distinguish between the type I and type II seesaw models of this kind. Clearly, there are two obvious avenues to explore in this connection.  One can look at the flavor structure of  the proton decay final states and secondly the predictions for rare lepton flavor violsting decays of type $B(\mu\to e+\gamma)$ and $B(\tau\to \mu (e) +\gamma)$. We investigate both these questions in this paper.

%
%Interestingly, this (approximate) equality can be realized
%in the light neutrino mass matrix using  naturalness argument~\cite{Akhmedov:1999uw}.
%
%However, the mixing angle and the squared mass difference ratio
%are not related in general.
%If there is (approximate) orthogonality of row vectors in the matrix, 
%a cancellation of the 13 mixing can happen.
%
%We therefore need a predictive picture of the light neutrino mass matrix where such an empirical relation between the mixing angle and the mass squared differences can arise naturally.
%
%One predictive  scenario is the grand unified theory (GUT).
%
%if the normal hierarchy of the neutrino masses are assumed.
%This is obtained from the naturalness of the light neutrino mass matrix.
%If there is (approximate) orthogonality of row (or column) of the matrix, 
%a cancellation of the 13 mixing can happen.
%Interestingly, this native naturalness relation can agree with the
%experimental measurements.
%Can the naive relation be written in a more predictive form?
%The purpose of this talk is to investigate the predictiveity
%of the 13 neutrino mixing in the SO(10) Grand Unified Theories (GUTs).
%
The reason to suspect that lepton flavor violation can probe different models is that in SUSY seesaw models all leptons are accompanied by their bosonic partners, the sleptons and their mixings can lead to large flavor violating effects at low energies. To prevent excessive flavor violating effects, one generally assumes that at some high scale, all slepton masses are equal. However as we extrapolate the theory down to the weak scale, slepton mixing are generated by the large neutrino mixings hidden in the Dirac Yukawa couplings of 
right-handed neutrinos \cite{Borzumati:1986qx}. In generic type I seesaw models without additional inputs, these rates can be suppressed by simply changing the seesaw scale since these mixings are proportional to products of $Y_\nu$ matrix elements and if the seesaw scale is lowered, seesaw formulae for neutrino masses demand that $Y_\nu$ become smaller in absolute magnitude thereby reducing the $\mu\to e+\gamma$ etc. rates. The same thing also happens in  SU(5) GUTs, where the size of the Dirac neutrino Yukawa coupling is a free parameter because the right-handed neutrino  is a gauge singlet and no definite prediction is possible.
	
However, in predictive GUT models such as the minimal renormalizable SO(10), coupling unification and fermion mass fits determine all Yukawa couplings of the theory and in particular, the Yukawa couplings of the RH neutrinos. Furthermore the values of the RH neutrino masses are also predicted making the lepton flavor violation predictions more definite. There are also separate contributions to LFV amplitudes coming from type II seesaw Yukawa couplings \cite{Joaquim:2006mn}, which are not generally considered. Another contribution to LFV  in all GUT models comes from charged fermion-quark-color triplet Higgs ($e^c u^c{H}_T$)  
coupling \cite{Barbieri:1994pv} which is independent of the seesaw mechanism;  it is also predicted in these models since fermion mass fits also determine those couplings. Moreover,
since the flavor structure of the Yukawa matrices is related to the
lepton flavor violations (LFV), and the proton decay amplitudes within a SUSY SO(10) framework, these two predictions are in principle connected.
We investigate this question in this paper for a specific realistic SO(10) model.
%Using the experimental constraints of them,
%one can obtain favorable Yukawa structure, which predicts the 
%neutrino mixing parameters. We investigate this connection between proton decay and lepton flavor violation in this paper.
 We note that although some early studies of LFV in SO(10) models have been carried out in Ref.\cite{Calibbi:2006nq}, where
 assumptions are made about expected relations between neutrino Yukawa matrix $Y_\nu$ and $Y_u$, in realistic models
there may be significant deviations from these relations. We find this to be true in models we are considering where in addition to new forms for the {\bf 10} couplings, there are  important $\overline{\bf 126}$ Yukawa contributions to flavor mixing.  

%The aim of this paper is to present predictions for LFV, proton decay, and neutrino mixing parameters
%in this class of realistic and predictive renormalizable SO(10)GUT models that invoke flavor structure to suppress proton decay. As noted above, in our previous paper, we presented the analysis with type II seesaw formula for neutrino masses \cite{Dutta:2004zh}. 
The new results of this paper are:
(i) we present a detailed analysis of the necessary texture to obtain ``large'' $\theta_{13}$ and suppressed proton decay for type I seesaw case and in particular discuss how the relation of 13 mixing and the ratio of the squared mass differences, Eq.(\ref{13-massratio}) emerges:
%(ii) we review how two large neutrino mixing angles and one small 13 mixing 
%are naturally obtained in the SO(10) model with $\overline{\bf 126}$ Higgs coupling,
%
(ii) we show that in the case of type II seesaw mechanism, 
 zero value of 13 neutrino mixing is
disfavored if we want to suppress proton decay via flavor structure;  (iii) we also show  that  the measured value of 13 neutrino mixing
prefers $\delta_{\rm PMNS} \sim \pi$ for the CP phase in the neutrino oscillation paramaters and (iv)  we compare the predictions of LFV for both type I and II seesaw and show how the required flavor structures in type I and type II seesaw  cases
can provide different predictions for LFV decays: in particular, we find that for the type I seesaw case, $B(\tau\to \mu +\gamma)$ can be observable while at the same time time suppressing $B(\mu\to e+\gamma)$, whereas in the type II seesaw case, $B(\tau\to \mu +\gamma)$ is always suppressed.
 
%
%In the case of type I seesaw, since
%the light neutrino mass matrix is a complicated function of the parameters compared to the type II case
%due to the inverse of the Majorana mass matrix, obtaining the Yukawa texture that leads to desired
% neutrino mass ratios and mixings naturally is nontrivial. It turns out that
%the required flavor structure is very different from the type II seesaw case discussed in \cite{Dutta:2004zh},
%andleads to suppression of proton decay operators somewhat more naturally.
%
%Moreover, because of no cancellation is required among the component in the Yukawa coupling matrices,
%the flavor structure in type I seesaw provides a predictive feature 
%for the partial lifetime of baryon number violating nucleon decays.
%

%\cite{Calibbi:2006nq}

This paper is organized as follows: in section II, we discuss the Yukawa matrices in SO(10) model; in section III, we discuss the suppression of nucleon decay amplitudes; in section IV, we discuss the LFV constraints; in section V, we discuss the flavor structures and predictions for type II seesaw scenario; in section VI we discuss the flavor structures and predictions for type I seesaw scenario; in section VII we discuss the predictions for lepton flavor violation in type I and II seesaw scenarios; in section VIII we discuss the predictions for nucleon decay in type I and II seesaw scenarios and we conclude in section IX.

%In this paper, we investigate a predictive SO(10) GUT model
%with renormalizable Yukawa couplings,
%and study how the neutrino mixing angle can be predictable
%from the proton decay constraints.
%The patterns of LFV can be related to the SO(10) breaking pattern,
%which depends on the type I and II seesaw to obtain light neutrino masses.

\section{Yukawa matrices for fermions in renormalizable SO(10)}

The Yukawa terms in the superpotential of the renormalizable SO(10) model 
involve the couplings of {\bf 16}-dimensional matter spinors $\psi_i$
with Higgs fields belonging to {\bf 10} (denoted by $H$) and  $\overline{\Delta}$, and $D$, representing the 
 $\overline{\bf 126}$, and {\bf 120} dimensional representations,
respectively and is given by: %The Yukawa superpotential is given by
\begin{equation}
W_Y = \frac12 \bar h_{ij} \psi_i \psi_j H + \frac12 \bar f_{ij} \psi_i \psi_j \overline{\Delta} 
+ \frac12 \bar h^\prime_{ij} \psi_i \psi_j D.
\end{equation}
%
%The Higgs fields, $H$, Later in the paper, we will see the the merits of selecting this representations.
%We will see the merits of this choice of the representations later.

%
This equation holds at the GUT scale. In order to write the effective Yukawa couplings below the GUT scale, we extract the effective $H_{u,d}$ fields of MSSM which are linear combinations of the Higgs doublet fields not only in $H, \overline{\Delta}, D$ fields but also in other fields e.g. $\Delta$ and {\bf 210} fields used to break the GUT symmetry while maintaining supersymmetry down to the weak scale.
We assume that the SO(10) symmetry is broken down to the standard model gauge symmetry
by the vacuum expectation values of $\Delta ({\bf 126}) + \overline \Delta(\overline{\bf 126})$ 
and the {\bf 210} Higgs field. Our conclusions below are independent of this i.e. the GUT symmetry could have been broken down by other fields such {\bf 54}, {\bf 45} etc as long as they do not contribute to fermion masses. 
We also assume that only one pair of the linear combinations of the Higgs doublets
remains massless (or more precisely, weak scale Higgsino mass, which is much smaller than
the GUT scale) to break the electroweak symmetry.
Using this, we can write the Yukawa matrices that give rise to the fermion masses as those
given by the linear combination of the original $\bar h$, $\bar f$ and $\bar h^\prime$
couplings:
\begin{eqnarray}
Y_u &=& h + r_2 f + r_3 h^\prime, \\
Y_d &=& r_1(h+f+h^\prime), \\
Y_e &=& r_1(h-3f+c_e h^\prime), \\
Y_\nu &=& h-3r_2f +c_\nu h^\prime, \label{Dirac-neutrino-Y}
\end{eqnarray}
where $r_i$ and $c_e$, $c_\nu$ are the functions of 
Higgs mixings, and $h$, $f$ and $h^\prime$ matrices
are the original couplings multiplied by Higgs mixings.
More details about these equations can be found in Ref.\cite{Dutta:2004zh}.

The $\overline{\bf 126}$ Higgs Yukawa coupling includes both
left- and right-handed Majorana neutrino couplings:
\begin{equation}
\psi \psi \overline\Delta \supset \ell \ell \overline\Delta_L
+ \bar\nu \bar\nu \overline\Delta_R,
\end{equation}
where $\overline\Delta_L$ denotes a SU(2)$_L$ triplet
in the $\overline{\bf126}$ representation.
As a result, 
in the case of triplet-part dominant type II seesaw neutrino mass \cite{Schechter:1980gr},
the neutrino mass is (approximately) proportional to the coupling matrix $f$.
It is clear from the above equation that there is an intimate connection between the lepton and the quark sector since the same Yukawa coupling matrix $f$ appears both in the charged fermion sector as well as the neutrino sector due to the seesaw mechanism. 
We assume CP conservation prior to symmetry breaking so that $h$ and $f$ are real symmetric matrices and
we can choose a basis so that $h$ is diagonal. The matrix $h^\prime$ is imaginary and anti-symmetric
so that the total number of coupling parameters in the theory (prior to any assumption about nucleon decay) is twelve. Combined with the other parameters
for the model i.e. (adding in $r_i, c_{e,\nu}$) and the $v_{BL}$ which gives the overall scale of neutrino masses)  the total number of parameters is eighteen. 
This is smaller than the number of observables i.e. 13 in the charged fermion sector and six in the neutrino sector excluding the Majorana phases. So the model 
is predictive. When combined with the flavor ansatz for suppressing nucleon decay mentioned above, the number becomes fifteen and the predictive power increases as we saw in Ref.\cite{Dutta:2004zh}.

\section{Suppression of nucleon decay amplitude}
In this section, we review the flavor suppression mechanism for nucleon decay in this model proposed in \cite{Dutta:2004zh}. The dominant proton decay in SUSY GUTs arise from the exchange of color triplet higgsinos which are part of the
{\bf 10}, $\overline{\bf 126}$ and {\bf 120} fields. The effective dimension five operator induced by them can be written as
\begin{equation}
-W_5 = \frac12 C_L^{ijkl} q_k q_l q_i \ell_j + C_R^{ijkl} e_k^c u^c_l u^c_i d^c_j.
\end{equation}
where the coefficients $C_{L,R}$ are functions of $h$, $f$ and $h^\prime$ couplings:
\begin{eqnarray}
&\!\!\!\!\!\!\!\!\!\!&\!\!\!\!\!\!\!\!\!
C_L^{ijkl}= %&\!\!\!\!=&\!\!\!\! 
\sum_a \frac1{M_{T_a}} (X_{a1} \bar h +X_{a4} \bar f + 
\sqrt2 X_{a3} \bar h^\prime)_{ij}
(Y_{a1} \bar h + Y_{a5} \bar f)_{kl}, \label{CL} \\
&\!\!\!\!\!\!\!\!\!&\!\!\!\!\!\!\!\!\! 
C_R^{ijkl}= %&\!\!\!\!=&\!\!\!\! 
\sum_a \frac1{M_{T_a}} (X_{a1} \bar h -X_{a4} \bar f + 
\sqrt2 X_{a2} \bar h^\prime)_{ij}
(Y_{a1} \bar h - (Y_{a5}-\sqrt2 Y_{a6}) \bar f + \sqrt2 (Y_{a3}-Y_{a2}) \bar h^\prime)_{kl}, \label{CR}
\end{eqnarray}
where $M_{T_a}$ are masses of colored Higgs triplets,
and $X$ and $Y$ are the colored Higgs mixings
(Details are given in Ref.\cite{Dutta:2004zh}).

%Why {\bf 120} Higgs coupling is preferred rather than {\bf 10}$^\prime$ Higgs?
%
%\begin{equation}
%C_L \sim (h+U_{15} f)_{ij}(h+V_{14} f+V_{12} h^\prime)_{kl}
%\end{equation}

As is noted in the introduction in order
to suppress the nucleon decay amplitudes,
one has to consider
\begin{enumerate}
\item
Choice of SUSY breaking parameter (heavy squarks being preferable).
\item
%Heavy 
The (lightest) 
colored Higgs triplet is heavy
and/or the colored Higgs mixings are small.\footnote{We investigate
what kind of SO(10) breaking vacua is preferable to obtain heavy colored Higgs
in the current scheme in Ref.\cite{Dutta:2007ai}.}
\item
Special structure of Yukawa matrices.
\end{enumerate}
%
% These three items may be cooperated to satisfy the
%current experimental bounds.
All the above requirements may work  in tandem  to satisfy the current experimental constraints on proton lifetime without invoking cancellation.
We will concentrate on the item 3 in this paper, and consider 
what kind of flavor structure is favorable to suppress
the nucleon decay operators, while at the same time giving correct prediction for the 
fermion masses and mixings.

%We are studying the case 3, where what kind of flavor structure is favorable to suppress
%the nucleon decay operators.
The important features that arise in the discussion of  suppressing nucleon decay operators 
by a flavor structure
are as follows~\cite{Dutta:2004zh,Dutta:2009ij}:
\begin{enumerate}

\item
In the current setup, there are multiple pairs of Higgs fields,
and thus, there is freedom to cancel the decay amplitude.
However, the required cancellations  in
both $C_L$ and $C_R$ operators are large, especially for large $\tan\beta$.
Sometimes,  $C_R$ operators are ignored
since they are suppressed by Higgsino dressing rather than gaugino dressing.
It is true that the contribution from $C_R$ to the nucleon decay amplitude is rather suppressed
compared to $C_L$.
However, in the case that the 1st generation masses are obtain
by a choice of $h$ and $f$ coupling and $h_{11}, f_{11} \sim y_d$ (down quark Yukawa coupling),
the $C_R$ operators are far from being small. %(unless $\tan\beta$ is very close to 1).
Indeed, even in the minimal-type of SU(5) model,
when only one pair of Higgs fields couple to fermions
and the dimension five operators are roughly in the form of $Y_d Y_u$,
 and leads to trouble due to the fact that both $C_L$ and $C_R$
 cannot be suppressed simultaneously from the flavor structure \cite{Dutta:2004zh}
 unless the colored Higgs spectrum is extended \cite{Dutta:2007ai}. 
%due to this issue (unless the choice 1 or 2 is applied).
%However, 
%
The SO(10) model (in which the cancellation is compatible with quark/lepton mass hierarchy)
provides the rough structure of proton decay operators given by $Y_d Y_d$,
and the size of $C_L$ and $C_R$ operators are much larger than the current experimental bound.
We also stress that in SO(10) models, the 
Higgs triplets in $\bf 10$ and $\overline{\bf 126}$ have opposite $D$-parity and
therefore appear with opposite signs in the $C_{L,R}$ expressions as in Eqs.(\ref{CL}) and (\ref{CR}).
Therefore, the cancellation (between $h$ and $f$) is unnatural. On top of this,
such  cancellation need to happen for each decay mode.

\item
Since the cancellation between $h$ and $f$ is not naturally realizable,
%the quark mass hierarchy,
it is preferable that $h$ coupling structure is similar to the up-type quark mass hierarchy,
and the down-type quark and charged
lepton mass hierarchy is generated by $f$ and $h^\prime$ matrices.
Even in this case, the $h_{ij} h_{kl}$ contribution is comparable to the current experimental bound 
even if $h_{11} \sim y_u$ and $h_{22} \sim y_c$,
where $y_u$ and $y_c$ are up and charm quark Yukawa couplings.
%
%
%
%
%Because the Higgs triplets in $\bf 10$ and $\overline{\bf 126}$ have
%opposite $D$-parity, there are opposite signs in the $C_{L,R}$ expressions,
%which makes the cancellation unnatural especially for large $\tan\beta$.
%
%If we do not adopt the rank 1 $h$ structure, 
%
%We emphasize that such cancellations are unnatural since Higgs triplets
%in $\bf 10$ and $\overline{\bf 126}$ have opposite $D$-parity and
%there are opposite signs in the $C_{L,R}$ expressions as in Eqs.(13) and (14).
%
Therefore to avoid unnatural cancellation which has to be implemented for each decay mode,
the simplest choice appears to be to have $h_{11} \ll y_u$ and $h_{22} \ll y_c$.
This implies that the $h$ coupling has no role in generating the 
1st and 2nd generation masses and only 3rd generation masses
are generated by $h$; in other words,
$h$ coupling can be chosen to be a rank 1 matrix to leading order.

%....($h$ is rank 1)

\item

If the $h$ coupling matrix is rank 1,
and 1st and 2nd generation masses
are generated from
$f$ and/or $h^\prime$ couplings, thus avoiding 
 the need for cancellation between $h$ and $f$.
We however still need suppression of the contributions from $f$ and $h^\prime$.
The simplest choice seems to be that
the 1st generation masses and Cabibbo mixing are generated
by $h^\prime_{12}$, and $f$ coupling does not contribute to them.
Since the $h^\prime$ matrix is anti-symmetric,
it does not contribute to the $kl$ part of the $C_L$ operator Eq.(\ref{CL}).
In fact, since the down quark Yukawa coupling is much too large to 
satisfy the current experimental bounds for the the nucleon decays,
 the question of how to generate the first generation masses and Cabibbo angle
becomes an important one and
this is the reason way we adopt a ${\bf 120}$ Higgs representation
instead of an extra {\bf 10} Higgs field.
%
%In order to generate the first generation masses and Cabibbo angle,
%One can employ extra $\bf 10$ Higgs field to generate them.
%However, in that case, the size of the extra $\bf 10$ Higgs coupling is quite harmful
%to the proton decay operators.
%The choice of $\bf 120$ Higgs field is favorable to generate the first generation masses
%since the $h^\prime$ matrix is antisymmetric,
%and it does not contribute to the $kl$ part of the $C_L^{ijkl}$ operator.

%In fact, the size of down-quark Yukawa coupling is harmful
%to satisfy the current experimental bounds,
%
%and the choice of {\bf 120} Higgs representation is helpful.

\item

The size of $f_{1i}$ components are also
important for a natural suppression of the proton decay amplitude.
If the $f$ coupling does not contribute to the down-quark and electron masses,
and Cabibbo mixing angle,
$f_{11}$ and $f_{12}$ can be taken small.
%
%
%We note that $f_{12}$ can be parametrized to be zero as a choice of basis,
%if $h$ is a rank 1 matrix.
%
%The $f_{11}$ component can be small to fit the fermion mass.
%
In fact, this choice of $f$ and $h^\prime$ coupling matrices
can be simply consistent to the empirical relations:
 $\sin\theta_C \simeq \sqrt{m_d/m_s}$,
and $m_d m_s m_b \simeq m_e m_\mu m_\tau$. 
Under the assumption that the $h$ coupling is dominantly large and rank one
and $h^\prime$ and $f$ are the correction needed to generate the 
masses of the first and second generation quarks and leptons, respectively,
the quark mixings $V_{cb}$ and $V_{ub}$
are generated to be small
due to the left-right symmetry.

\item

The last piece in this discussion involves the magnitude of the 13 element of $f$ coupling.
In the case of type II seesaw (triplet part dominant),
a sizable $f_{13}$ is needed to obtain a proper solar mixing and
neutrino mass ratio.
In this case, one still needs a cancellation in the nucleon decay amplitudes.
%
%The size of $f_{13}$ is one of the key,
%and 
It is important to study
whether the $f_{13}$ coupling can be made to be small.
%for the proton decay suppression.
%
The size of $f_{13}$ is important not only for the nucleon decay suppression
but also for the lepton flavor violation
as we will see next section.

%On the other hand, the $f_{13}$ component is needed
%for type II seesaw fit since the neutrino mass matrix is proportional to $f$.
%
%Therefore, one still needs a cancellation.
%
%The size of $f_{13}$ is one of the key,
%It is one of the issue
%whether the $f_{13}$ can be made to be small
%for the proton decay suppression.
%
%The size of $f_{13}$ is also important for the lepton flavor violation
%as we will see next section.

\end{enumerate}

%
%We will discuss how the neutrino mixing angles 
% (two large and one small mixings)
%are obtained naturally.

\section{Lepton Flavor Violation}
We now turn to the discussion of lepton flavor violating rare decays in this class of models. It is well known that
the SUSY GUTs predict the lepton flavor violating rare decays due to the possibility of large flavor mixings in the superpartner sector.
In order to avoid too much FCNCs induced by them, the flavor universality of the 
SUSY breaking mass parameters is often invoked.
Even so, the loop correction via the heavy GUT particles 
can generate  flavor violation.
Though there can be new flavor changing effects in quark sector in SUSY GUTs,
the recent LHC result imply no large effects  \cite{Amhis:2012bh,:2012ct}.
We, therefore, concentrate on the flavor violation in the lepton sector.

The typical sources of LFV in SUSY SO(10) are as follows:
\begin{enumerate}

\item
$e^c u^c H_C$ coupling (which is $\bf{10}_m \cdot \bf{10}_m \cdot H_5$ coupling in SU(5)).

\item
Dirac neutrino coupling ($Y_\nu\ell \nu^c H_u$).

\item
Majorana neutrino coupling ($f_\nu \ell \ell \overline\Delta_L$).

\end{enumerate}
In SO(10) model, the Dirac neutrino Yukawa coupling
$Y_\nu$ is the effective combination of {\bf 10},  $\overline{\bf 126}$ and {\bf 120} 
doublet coupling with appropriate mixing ratios (as in Eq. 6)  whereas the Majorana neutrino coupling
$f_\nu$ is unified to the $\overline{\bf 126}$ coupling (up to a Clebsch-Gordon coefficient).

One can express 
the RGE-induced off-diagonal elements of SUSY breaking masses
%the parameters $M^2_{12}$ and $M^2_{13}$ can also be written
 in terms of the $Y_\nu$ and $f_\nu$ Yukawa couplings as follows:
\begin{eqnarray}
\Delta M^2_{ij}\simeq -\frac{3m^2_0+A^2_0}{16\pi^2}\sum_k (Y^*_{\nu, ik}Y_{\nu,kj}) \ln\frac{M^2_U}{M^2_{R_k}}
-\frac{9m^2_0+3A^2_0}{16\pi^2} \sum_k (f^*_{\nu, ik}f_{\nu,kj}) \ln\frac{M^2_U}{M^2_\Delta},
\end{eqnarray}
The initial flavor universality is assumed and $M_{ij}^2 = m_0^2 {\bf 1}$,
and $A_0$ stands for an universal trilinear scalar coupling.

In general, the off-diagonal elements by these couplings can be parametrized as
%{\bf I SUGGEST WE WRITE $\Delta M^2_{IJ}$ IN TERMS OF $Y_\nu$ AND F-MATRICES SO THAT WE CAN DIRECTLY CONNECT TO THEORY; THE ANGLES $\Theta_{ij}$ BELOW ARE NOT NEUTRINO MIXINGS}
\begin{equation}
\Delta M_{ij}^2 \sim -\kappa m_0^2 U 
	\left(
		\begin{array}{ccc}
		    k_1 &&\\
		    & k_2 & \\
		    && 1
		\end{array} 
  	\right) U^\dagger,
\label{off-diagonal}	
\end{equation}
where the unitary matrix $U$ is just for the parameterization in general.
The mixing angles in $U$ depends on the source of flavor violation.
Once the source of the flavor violation is specified (with some assumptions),
numerical quantities in the parameterization can be related to the observed values.
For example, if the source only comes from the Dirac neutrino Yukawa coupling,
$Y_\nu \equiv U_L Y_\nu^{\rm diag} U_R^T$ 
(in the basis where charged-lepton mass matrix and the right-handed Majorana mass matrix are diagonal),
we obtain $U = U_L^*$,
$\kappa = (3+m_0^2/A_0^2)/(8\pi^2) (Y_\nu^{\rm diag})_{33}^2 \ln M_U/M_{R_3}$.
The quantities $k_i$ are specified by the hierarchy of the eigenvalues
of the Yukawa couplings (with RGE corrections).
If $U_R = {\bf 1}$ is supposed, the unitary matrix $U_L^*$ corresponds to the neutrino mixing matrix
(up to renormalization group evolution).
In general, the unitary matrix $U_L$ is different from the neutrino mixing matrix.
However, it is often assumed that the large neutrino mixing angles originates from $U_L$.
In the SO(10) model, the angles in $U$ in the parameterization is different from the neutrino mixing angles,
and those are computed from the expression of $Y_\nu$.
In the triplet part dominant type II seesaw (i.e. type I seesaw part is negligible),
the unitary matrix $U$ corresponds to the neutrino mixing matrix 
if the $f_\nu$ coupling is the dominant source of the flavor violation.

%The numerical quantity $\kappa$ is specified by the size of the source and the 
%log factor from RGE, and the ratio of scalar mass $m_0$ and scalar trilinear coupling $A_0$.
%The quantities $k_i$ are specified by the hierarchy of the eigenvalues
%of the sources (with RGE corrections).
%The unitary matrix $U$ depend on the type of seesaw scenario
%as well as the sources of LFV.

The most stringent constraint is provided by the experimental upper limit on the branching 
ratio of $\mu \to e\gamma$ \cite{Adam:2011ch}:
\begin{equation}
{\rm Br} (\mu\to e\gamma) < 2.4 \times 10^{-12}.
\end{equation}
To suppress the $\mu\to e\gamma$ amplitude in SUSY models,
the 12 and 13 elements of the slepton mass matrices, given by the equation below, have to be small:
\begin{eqnarray}
M^2_{12} &=& - \kappa m_0^2 (\frac12 k_2 \sin2\theta_{12} \cos\theta_{23} + e^{i\delta} \sin\theta_{13}\sin\theta_{23})e^{i(\beta-\alpha)}, \\
M^2_{13} &=& - \kappa m_0^2 (\frac12 k_2 \sin2\theta_{12} \sin\theta_{23} - e^{i\delta} \sin\theta_{13}\cos\theta_{23})e^{i\beta},
\end{eqnarray}
where $\theta_{ij}$ denotes mixing angles in the matrix $U$ and $\alpha$, $\beta$, $\delta$ are phases in $U$.

The left-handed $\mu\to e\gamma$ amplitude can be cancelled by a choice of $k_2$ and mixing angles.
However, such cancellation cannot happen for both $\mu\to e\gamma$ and $\mu$-$e$ conversion
simultaneously.
As a consequence,
the suppression of $\mu \to e\gamma$ and $\mu$-$e$ conversion
requires (at least) one of the followings:
\begin{enumerate}

\item small $\kappa$ (which means that the relevant Yukawa couplings are small), 

\item both 12 and 13 mixings in $U$ are small, 

\item $k_i \ll 1$ (which means that the eigenvalues of Yukawa matrix is hierarchical) and 13 mixing is small.

\end{enumerate}

If $\kappa$ is small, $\tau\to \mu(e)\gamma$ is also small.
In the other cases ( (2) and (3) above), $\tau\to \mu\gamma$ can be sizable, staying just below the current experimental bound
while satisfying the bounds from $\mu\to e\gamma$ and $\mu$-$e$ conversion
if 23 mixing angle in $U$ of the source is large.
It is therefore important to sort out the models
to see if $\tau \to \mu(e) \gamma$ can be observed just below the current bounds,
with a potential for $\mu\to e\gamma$ to be observed near future.
We will study the patterns of LFV in our SO(10) model using type I and type II seesaw models once we fit the neutrino masses and suppress proton decay.

%In our model of rank 1 $h$ coupling with type II seesaw,
%the situation should be the case 1 for the Majorana coupling origin,
%and the case 2 and 3 for Dirac neutrino coupling origin.
%Since the 23 mixing angle in the Dirac neutrino coupling is small,
%$\tau \to \mu\gamma$ prediction is much below the current experimental bound.
%The branching ratio of $\mu\to e\gamma$ can be in the range to be observed
%at the decay experiments.

\section{Predictions for flavor structures in type II seesaw}

%\subsection{Simple realization of neutrino mixings}

We discuss the suitable choice of flavor structure in lepton sector,
and how to accommodate 
two large neutrino mixing angles(for atmospheric and solar neutrino oscillations)
and one (relatively) small 13 neutrino mixing angle 
 naturally in type II seesaw in SO(10) GUT \cite{Dutta:2009ij}.

Let us first describe the flavor structure in a general case. 
We start from a basis where neutrino mass matrix is diagonal.
In this basis, the neutrino mixing matrix is equal to the diagonalization matrix of 
charged lepton mass matrix $M_e$.
\begin{equation}
U^*_{\rm PMNS} M_e M_e^\dagger U_{\rm PMNS}^T = (M_e^{\rm diag})^2.
\end{equation}
%
%(As a convention, $U^*$ is the PMNS matrix).
%
Since the muon mass is much smaller than the tau lepton mass,
one can decompose $M_e$ as
\begin{equation}
M_e = M_e^0 + \delta M_e,
\end{equation}
where $M_e^0$ is a rank 1 matrix and generate tau mass,
and elements of $\delta M_e$ is smaller than $M_e^0$. 
The rank 1 matrix is written in general as
\begin{equation}
M_e^0 = \left(
 \begin{array}{c}
  c \\
  b \\
  a 
  \end{array}
\right)
\left(
 \begin{array}{ccc}
  c &
  b &
  a 
  \end{array}
\right).
\end{equation}
In general we can make the rank 1 matrix symmetric by rotating the right-handed lepton fields,
and therefore, we write it in the symmetric form.
In that basis, $\delta M_e$ is not necessarily a symmetric matrix.
The rank 1 matrix can be diagonalized by two angles (or one can say one of the three angles is unphysical
if $\delta M_e = 0$).
A unitary matrix to diagonalize the rank 1 matrix can be written as
\begin{eqnarray}
U_0 &=& 
\left(
 \begin{array}{ccc}
  1 & 0 & 0 \\
  0 & \cos\theta_a & -\sin\theta_a \\
  0 & \sin\theta_a & \cos\theta_a
 \end{array}
\right)
\left(
 \begin{array}{ccc}
  \cos\theta_s & -\sin\theta_s & 0 \\
  \sin\theta_s & \cos\theta_s & 0 \\
  0 & 0 & 1
 \end{array}
\right) \\
&=&
\left(
 \begin{array}{ccc}
  \cos\theta_s & - \sin\theta_s & 0 \\
  \cos\theta_a \sin\theta_s & \cos\theta_a \cos\theta_s & - \sin\theta_a \\
  \sin\theta_a \sin\theta_s & \sin\theta_a \cos\theta_s & \cos\theta_a
 \end{array}
\right),
\end{eqnarray}
where 
\begin{equation}
\tan\theta_s = \frac{c}{b}, \qquad
\tan\theta_a = \frac{\sqrt{b^2+c^2}}{a}.
\end{equation}
%
%The 13 element of $U_0$ is zero.
%
We define a basis rotated by $U_0$.
(We attach ``hat" to distinguish from the original one.)
\begin{equation}
\hat M_e = \hat M_e^0 + \delta \hat M_e,
\end{equation}
where $\hat M_e^0 = {\rm diag} (0,0,m_3)$ ($m_3 = a^2+b^2+c^2$),
and $\delta \hat M_e = U_0 (\delta M_e) U_0^T$. 
Define the diagonalization matrix of $\hat M_e$ as $V_e$.
Then the mixing unitary matrix is written as
\begin{equation}
U_{\rm PMNS} = V_e U_0.
\end{equation}
The 13 element of $U_{\rm PMNS}$ is
\begin{equation}
U_{e3} = -(V_e)_{12} \sin\theta_a + (V_e)_{13} \cos\theta_a, 
\end{equation}
%where $\gamma$ is a phase.
%(The relation of this $\gamma$ differs from the leptonic CP phase which appear in the neutrino oscillation.)
%
From our  assumption $m_3 \gg (\delta M_e)_{ij}$,
%$\hat M_e$ is hierarchical, 
%and 
we expect $(V_e)_{13}$ and $(V_e)_{23}$  to be small.
%
%the off-diagonal elements are expected to be small.
%
The angle $\theta_a$ is almost same as the 23 mixing (for atmospheric neutrino oscillations)
if $(V_e)_{23}$ is tiny.
The solar angle is modified from $\theta_s$ by $(V_e)_{12}$.
Naively, if $(V_e)_{13}$ can be negligible, 
the 13 mixing is $(V_e)_{12}/\sin\theta_{\rm atm}$.
At this stage, this is just a parametrization of 13 mixing.
But it is very useful to work using the unification picture.
Interestingly, the experimental measurements of the 13 mixing are
 consistent with $(V_e)_{12} = V_{us}$, which is the Cabibbo angle in the quark sector.

We note that 
$U_{e3}$ is exactly equal to zero
in the limit where only 33 element of $\delta M_e$ is non-zero
(in this limit, the rank of $M_e$ is 2, and electron is massless).
In general, $\theta_a$ and $\theta_s$ are large.
If $(\delta M_e)_{33}$ dominantly generate muon mass,
the 13 mixing angle is naturally small 
compared to the others,
and the naive size of the 13 mixing is expected to be electron/muon mass ratio.

Generically, there is no reason why
$(\delta M_e)_{33}$ is dominant compared to the other elements
and $(V_e)_{12}$ is small (in the ``hat" basis).
%
%This is nice feature in type II seesaw,
%where $\delta Y_e$ comes from $\overline{\bf 126}$ Higgs coupling
%and proportional to neutrino  matrix.
%
In the triplet-dominant type II seesaw, however,
such situation is quite natural
because 
the correction of the charged lepton mass matrix
and the neutrino mass matrix can be unified (up to factor)
to the $\overline{\bf 126}$ Higgs coupling $f$.
Besides, the $\overline{\bf 126}$ coupling is good to generate
the strange quark and muon masses for the Clebsh-Gordon coefficient.
We emphasize that the situation
is consistent with the suppression of nucleon decay amplitudes
by a flavor structure.
%
%In this section, we will briefly review the almost rank 1 structure
%to realize successful fermion masses and mixings \cite{Dutta:2009ij}.
%
%Brief review of rank 1 structure.
%{\bf 126} Higgs representation has a merit of natural
%realization of fermion masses and mixings in the rank 1 structure.
%
As we have explained,
we are studying the situation where the fermion masses are dominantly given
by rank 1 matrix $(h)$, which gives third generation charged-fermion masses,
and the $f$ and $h^\prime$ coupling matrices give the first and second generation
masses, and fermion mixings.
Under this assumption, qualitative structure of the fermion masses and mixings can be 
easily reproduced in the case of triplet-part dominant type II seesaw.
In addition to the natural realization of two large and one small neutrino mixings,
 the quark mixings are small under this assumption due to left-right
symmetry in  SO(10).

%
%The neutrino mixing angles are generically large, because they are roughly the angles in the
%relative diagonalization unitary matrices of $h$ and $f$.
%However, relating to the fact that there are only two physical diagonalization angles of the rank 1 matrix,
%the 13 mixing angle can be small, while two other angles, corresponding to the solar and atmospheric
%mixing angles, are large in general.
%

Let us illustrate the feature of the flavor structure in triplet-dominant type II seesaw.
We choose a basis where $f$ matrix is diagonal, and parametrize
\begin{equation}
h = h_{33}\left(
 \begin{array}{ccc}
  c^2 & b c & a c \\
  b c & b^2 & a b \\
  a c & a b & a^2
 \end{array}
\right),
\qquad
 f =  f_{33}
\left(
 \begin{array}{ccc}
  \lambda_1 & 0 & 0 \\
  0 & \lambda_2 & 0 \\
  0 & 0 & 1
 \end{array}
\right).
\end{equation}
In this basis, the neutrino mass matrix is (nearly) diagonal
neglecting the type I seesaw term (which is assumed to be suppressed by heavy right-handed neutrinos).
The neutrino mixing matrix is the diagonalization matrix of $Y_e$.
For simplicity, the $h^\prime$ contribution is neglected in this illustration.
Then, in the limits of $\lambda_1, \lambda_2 \to 0$ and $f_{33} \ll h_{33}$,
one obtains
\begin{equation}
\tan^2\theta_{\rm atm} = \frac{b^2+c^2}{a^2},
\qquad
\tan\theta_{\rm sol} = \frac{c}{b},
\qquad
\sin \theta_{13} = 0.
\end{equation}
The limit $\lambda_1, \lambda_2 \to 0$
corresponds to the massless electron limit
and two zero eigenvalues of neutrinos.
Therefore, it is naturally realized that two solar and atmospheric mixings are large,
and small $\sin\theta_{13}$.
The size of the 13 mixing is expected to arise from the electron/muon mass ratio,
as well as the ratio of mass squared differences, $\Delta m^2_{12}/\Delta m^2_{23}$.
In order to fit the electron mass without fine-tuning, 
the $h^\prime_{12}$ component is useful.
If the fine-tuning to fit the electron mass is absent\footnote{
If one allows a cancellation to fit the electron mass,
the prediction is lost, and the 13 mixing can have O(1) factor ambiguity.
However, if one fit all the fermion masses and mixings without $\bf 120$ 
Higgs coupling (only $\bf 10$ and $\overline{\bf 126}$ couple to fermions \cite{Babu:1992ia,Matsuda:2000zp}), 
there is predictivity of 13 mixing due to the reduction of 
parameters \cite{Fukuyama:2002ch}.
In this case, 
the electron mass is obtained by a cancellation.
The prediction of the 13 mixing is related to the Cabibbo mixing angle
%and 
(roughly $V_{us}/\sqrt2$),
and the prediction can agree with the recent measured value.
%as it is mentioned in Babu's talk \cite{Babu}.
},
the 13 neutrino mixing can be directly related to the electron/muon mass ratio,
and $\Delta m^2_{12}/\Delta m^2_{23}$.

The first possibility is that the 13 neutrino mixing only depends on 
electron/muon mass mass ratio.
This can be constructed to make the $\overline{\bf 126}$ coupling matrix tri-bimaximal form
by using a discrete flavor symmetry \cite{Dutta:2009ij}.
In this case, 
the $\Delta m^2_{12}/\Delta m^2_{23}$ dependence can be dropped, and
the predicted 13 mixing is
\begin{equation}
\sin\theta_{13} \sim \sqrt{\frac{m_e}{m_\mu}} \sin \theta_{\rm atm} 
\sim \frac13 V_{us} \sin\theta_{\rm atm} \simeq 0.05.
\end{equation}
This does not match the measured 13 mixing angle.
We note that
the tri-bimaximal form of $\overline{\bf 126}$ is not preferable for the
nucleon decay suppression,
and we discard this possibility even without using the $\theta_{13}$ mixing.

The second possibility is that the 13 mixing angle is related to
both the electron/muon mass ratio
and $\Delta m^2_{12}/\Delta m^2_{23}$. 
The $\Delta m^2_{12}/\Delta m^2_{23}$ dependence of the 13 mixing angle
is related to the $f$ coupling structure,
which is related to the proton decay suppression, which we will see later.
Indeed, in the rank 1 structure, the size of $f_{11}$ and $f_{13}$
are important for both proton decay and LFV.
The small $f_{11}$ component is favorable to suppress proton decay amplitude,
and it allows prediction of the 13 neutrino mixing angle. 
%We note that the $f_{12}$ component can be parametrized to be zero without loss of generality
%if the $h$ coupling is assumed to be rank 1 matrix.

%In this rank 1 $h$ structure,
%the $h$, $f$ and $h^\prime$ coupling matrices are given as
%\begin{eqnarray}
%h = {\rm diag}. (0,0,h_3), \qquad
%f \sim 
%\left(
% \begin{array}{ccc}
%  0 & 0 & \lambda^3 \\
%  0 & \lambda^2 & \lambda^2 \\
%  \lambda^3 & \lambda^2 & \lambda^2
% \end{array}
%\right), 
%\qquad
%h^\prime \sim 
%\left(
% \begin{array}{ccc}
%  0 & \lambda^3 & \lambda^3 \\
%  -\lambda^3 & 0 & \lambda^2 \\
%  -\lambda^3 & -\lambda^2 & 0
% \end{array}
%\right),
%\end{eqnarray}
%
%where $\lambda\sim 0.2$.
%The smallness of the $f_{11}$ component is also consistent with
%the empirical relation $V_{us} \simeq \sqrt{m_d/m_s}$.
%
%Roughly saying,
%the 3rd generation masses are given by $h_3$,
%and the 2nd generation masses are given by $f_{22}$
%with $r_2$ and Clebsch-Gordon coefficient.
%The $h_{12}^\prime$ elements gives the first generation masses
%as well as the Cabibbo mixing angle.
%The absence of $f_{11}$ provides natural understanding of the
%so-called Goergi-Jarskog relation : $m_e m_\mu m_\tau \simeq m_d m_s m_b$
%by a choice of $c_e$.
%Other parameters can be chosen to be consistent with the other quark mixings and
%atmospheric and solar mixing angles.

%\subsection{Prediction in Type II (triplet dominant) seesaw}

%In this section, we will work 

Let us study
 how the 13 neutrino mixing angle is numerically restricted
in the rank 1 $h$ Yukawa structure:
%The rank 1 structure :
\begin{eqnarray}
h = 
\left(
\begin{array}{ccc}
0 & 0 & 0 \\
0 & 0 & 0 \\
0 & 0 & h_3
\end{array}
\right), \qquad
f = 
\left(
\begin{array}{ccc}
u & 0 & x \\
0 & y & z \\
x & z & w 
\end{array}
\right), \qquad
h^\prime = 
\left(
\begin{array}{ccc}
0 & c_1 & -c_2 \\
-c_1 & 0 & c_3 \\
c_2 & -c_3 & 0 
\end{array}
\right).
\end{eqnarray}
%
%$h^\prime$ coupling is needed to fit masses and mixings accurately.
%But we do not specify the Higgs representation to give the correction.
%
Note that we can parametrize $f_{12} = f_{21} =0$
without loss of generality, by diagonalizing 1st-2nd block of the matrix $f$.
As we have noted,
$u\to 0$ is preferable
to suppress proton decay amplitude and to obtain the empirical relation $V_{us} \simeq \sqrt{m_d/m_s}$.
%and Goergi-Jarskog relation without cancellation.
%
Roughly speaking,
the 3rd generation masses are given by $h_3$,
and the 2nd generation masses are given by $f_{22}$
with $r_2$ and Clebsch-Gordon coefficient.
The $h_{12}^\prime$ elements gives the first generation masses
as well as the Cabibbo mixing angle.
The absence of $f_{11}$ provides natural understanding of the
so-called Georgi-Jarskog relation (without a cancellation): $m_e m_\mu m_\tau \simeq m_d m_s m_b$
by a choice of $|c_e| \simeq 1$.
Other parameters can be chosen to be consistent with the other quark mixings and
atmospheric and solar mixing angles.

%If $u \to 0$, 
{}From $f_{12} = 0$,
we obtain a simple relation among the eigenvalues of $f$ and mixing angles:
%\footnote{
%Exchange of $f_1$ and $f_2$ corresponds to $\theta_{12} \to \theta_{12} + \pi/2$.
%}:
%
\begin{eqnarray}
\frac{f_1}{f_3} &=& u \sec^2 \theta_{13}^\nu  
         -e^{-2i\delta} \tan^2\theta_{13}^\nu
                   + e^{-i\delta} (1- u e^{2i\delta}) \sec\theta_{13}^\nu \tan\theta_{13}^\nu 
                                \tan\theta_{12}^\nu \tan\theta_{23}^\nu,  \\
\frac{f_2}{f_3} &=& u \sec^2 \theta_{13}^\nu
         -e^{-2i\delta} \tan^2\theta_{13}^\nu
                   - e^{-i\delta} (1- u e^{2i\delta})\sec\theta_{13}^\nu \tan\theta_{13}^\nu 
                                \cot\theta_{12}^\nu \tan\theta_{23}^\nu,
\end{eqnarray}
where
\begin{equation}
f = U_\nu^T 
 \left(
 \begin{array}{ccc}
 f_1 & & \\
 & f_2 & \\
 & & f_3 
 \end{array}
 \right) U_\nu,
\end{equation}
\begin{equation}
U_\nu=
\left(
\begin{array}{ccc}
\cos\theta^\nu_{12} & -\sin\theta_{12}^\nu & 0 \\
\sin\theta_{12}^\nu & \cos\theta_{12}^\nu & 0 \\
0 & 0 & 1
\end{array}
\right)
\left(
\begin{array}{ccc}
\cos\theta_{13}^\nu & 0 & -e^{i\delta} \sin\theta_{13}^\nu \\
0 & 1 & 0 \\
e^{-i\delta}\sin\theta_{13}^\nu & 0 & \cos\theta_{13}^\nu
\end{array}
\right)
\left(
\begin{array}{ccc}
1 & 0 & 0 \\
0 & \cos\theta_{23}^\nu & -\sin\theta_{23}^\nu \\
0 & \sin\theta_{23}^\nu & \cos\theta_{23}^\nu
\end{array}
\right).
\end{equation}
\begin{figure}[tbp]
 \center
  \includegraphics[width=10cm]{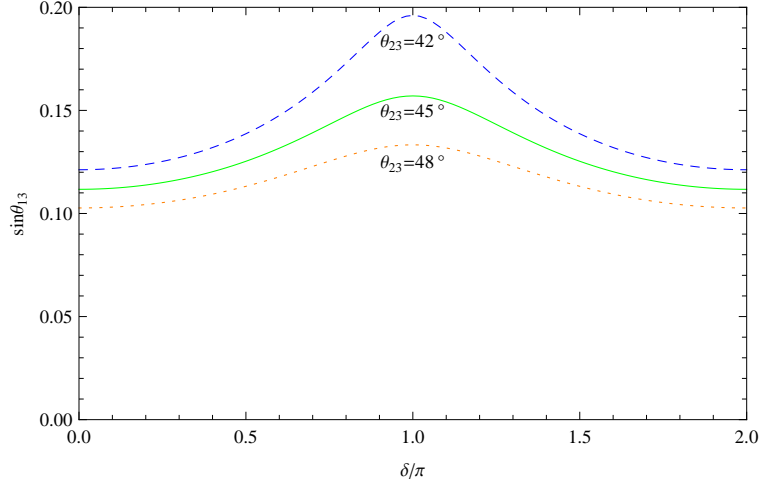}
 \caption{
 Magnitude of $\sin\theta_{13}$ depending on $\delta/\pi$ (and also 
 $\theta_{23}$, each choice of which gives one line), 
 which is restricted in a narrow band. 
}
\label{Fig1}
\end{figure}
In triplet-part dominant type II seesaw (and in normal hierarchy), we have
\begin{equation}
\frac{\Delta m^2_{12}}{\Delta m^2_{23}}
= \frac{|f_2|^2 - |f_1|^2}{|f_3|^2 - |f_2|^2}.
\end{equation}
In the limit $u\to 0$,
we obtain
\begin{eqnarray}
\frac{\Delta m^2_{12}}{\Delta m^2_{23}}
=
\frac{
4 \sin^2\theta_{13}^\nu \csc 2\theta_{12}^\nu \tan\theta_{23}^\nu
(\sin\theta_{13}^\nu \cos\delta + \cot 2\theta_{12}^\nu \tan\theta_{23}^\nu)}{
1- \sin^2\theta_{13}^\nu(2+ \cot^2\theta_{12}^\nu \tan^2\theta_{23}^\nu)
 - 2 \sin^3\theta_{13}^\nu \cos\delta \cot\theta_{12}^\nu
\tan\theta_{23}^\nu
}.
\end{eqnarray}
As a result,
$\theta_{13}$ is bounded from below by ${\Delta m^2_{12}}/{\Delta m^2_{23}}$.
Even if $u \neq 0$, 
$\theta_{13}$ has to be bounded %\footnote{Numerical analysis will be done later.}
because 
we obtain $f_1/f_3 \to u$, $f_2/f_3 \to u$
in the limit of $\theta_{13}\to 0$.
In figure 1, we plot the 13 mixing angle as a function of the CP phase $\delta$.
Since the 13 mixing angle is sensitive to the deviation from the 45 degree of 23 mixing,
we vary the 23 mixing angles.
It is not sensitive to the 12 mixing angle.
It is interesting to note that 
$\delta \sim \pi$ is preferable 
if the 13 mixing is larger value (depending on $\theta_{23}$).

The 13 neutrino mixing angle can be modified by the diagonalization matrix of charged lepton masses,
but the size correction is less than 0.05 radians\footnote{
The correction depends on relative phase freedom.
The maximal correction ($\sim 0.05$ radians ) occurs when the relative phase is 0 or $\pi$.}.
In conclusion, 
tiny $\theta_{13} (<0.05)$
is not allowed for the rank-1 structure with type II seesaw,
irrespective of the detail of fitting of fermion masses and mixings.
This is in contrast with the 
(nearly) tri-bimaximal model,
where $\theta_{13} \simeq 1/3 V_{us} \sin\theta_{23} \sim 0.05$.

%(proton decay. $f_{13}$ contribution is large and we should adjust the colored Higgs mixing parameters. )

%(LFV.
%$\tau\to \mu\gamma$ should be small (much below the current bound) due to 
%the $\mu \to e\gamma$ bound. 
%)

The suppression is 11,12 elements of $f$ coupling reduces the $ff$ contribution
of the nucleon decay amplitudes drastically.
Because the $f$ matrix is generated the large neutrino mixings and
the neutrino masses directly in the triplet dominant type II seesaw,
the 13 element of $f$ cannot be zero.
Therefore, 
it can still contribute to the nucleon decay amplitudes
and we need a cancellation which can be achieved by choosing the colored Higgs mixings to suppress them.

As we have noted, the LFV can be also generated from the $f$ coupling
if the doublet Higgs mixing angle is small and the original $\bar f$ coupling in the superpotential is large 
since (at least) SU(2)$_L$ triplet is lighter than the GUT scale.
In that case, however, 
$\mu\to e\gamma$ is generated and the doublet Higgs  mixing at GUT scale is bounded.
If the doublet Higgs mixing is O(1),
and $\bar f \sim f$, the size of Br($\mu\to e\gamma$) is the same order of 
the current experimental bound.
Since the size of 13 element is predicted in the type II seesaw model,
the Br($\tau\to\mu\gamma$) can be predicted, and it is about $O(10^{-10})$,% \cite{},
which is below the sensitivity at the LHCb.
In the type II seesaw neutrino, it is not possible
to enhance  
Br($\tau\to\mu\gamma$) to be of the order of the current experimental bound while
satisfying the bound of $\mu\to e\gamma$ and $\mu$-$e$ conversion.

\section{Type I seesaw}

In the case of triplet-part dominant type II seesaw,
the simple formula of the mixing angles are obtained,
and the two large and one small neutrino mixing angles can be easily realized.
On the other hand, for the type I seesaw case, the situation is more complicated
since the seesaw formula includes an inverse of the $f$ matrix.
In this section, we investigate how the fermion masses and mixings
are reproduced in type I seesaw with suppressed nucleon decays.

%How about the case of the type I?
%In fact, fitting the masses and mixings can be done, but those are complicate.
%Can a simple realization be obtained naturally similarly to the type II case?
%
%In a certain assumption, one can obtain a simple realization in the type I seesaw case.

Because the Dirac neutrino Yukawa coupling is given as
\begin{equation}
Y_\nu = h - 3r_2 f + c_\nu h^\prime,
\end{equation}
the type I seesaw neutrino mass matrix is proportional to 
\begin{equation}
Y_\nu f^{-1} Y_\nu^T
= (h+c_\nu h^\prime)f^{-1}(h-c_\nu h^\prime) - 6 r_2 h + 9 r_2^2 f.
\end{equation}
%and
%\begin{equation}
%h f^{-1} h 
% + c_\nu (h^\prime f^{-1} h - h f^{-1} h^\prime)
% - c_\nu^2 h^\prime f^{-1} h^\prime.
%\end{equation}
%\subsection{}
%
Because the up-type quarks are more hierarchical than down-type quarks,
$r_2$ %and the components of $f$ coupling  
is small
in the current scheme, 
and the last two terms can be negligible.
We, therefore, concentrate on the first term $N \equiv (h+c_\nu h^\prime) f^{-1} (h-c_\nu h^\prime)$.

We denote $h$ and $h^\prime$ matrices as
\begin{equation}
h = \left( 
 \begin{array}{ccc}
  h_1 && \\
  & h_2 & \\
  && h_3
 \end{array}
\right),
\qquad
h^\prime = \left(
 \begin{array}{ccc}
  0 & c_1 & -c_2 \\
  -c_1 & 0 & c_3 \\
  c_2 & -c_3 & 0
 \end{array}
 \right).
\end{equation}
We study whether a hierarchical $f$ matrix required by charged fermion fits and also proton 
decay suppression is consistent with the profile of observed neutrino mixings.
%We first study the property independent of how the neutrino mixings are obtained.
For this purpose, we first obtain the $f$ coupling by solving $N = (h+c_\nu h^\prime) f^{-1} (h-c_\nu h^\prime)$.
Denoting 
\begin{equation}
N = U {\rm diag}. (n_1,n_2,n_3) U^T,
\end{equation}
we obtain
\begin{equation}
f = (h-c_\nu h^\prime) U^* {\rm diag}.\left(\frac1{n_1}, \frac1{n_2}, \frac1{n_3}\right) U^\dagger (h+c_\nu h^\prime).
\label{solve-f}
\end{equation}
Defining ${\bf x}_i$ as
\begin{equation}
\left(
 \begin{array}{c}
 {\bf x}_1 \\
 {\bf x}_2 \\
 {\bf x}_3 
 \end{array}
\right)
= U^\dagger (h+c_\nu h^\prime),
\end{equation}
we obtain 
\begin{equation}
f = \frac{1}{n_1} {\bf x}_1^T {\bf x}_1
+
\frac{1}{n_2} {\bf x}_2^T {\bf x}_2
+
\frac{1}{n_3} {\bf x}_3^T {\bf x}_3.
\end{equation}
The neutrino mass matrix ${\cal M}_\nu$ is expressed as
\begin{equation}
{\cal M}_\nu = - N \frac{v_u^2}{v_R},
\end{equation}
where $v_u$ is a VEV of up-type Higgs field,
and $v_R$ is a VEV of {\bf 126} Higgs field
which breaks SO(10) down to SU(5).
(More precisely, this $f$ coupling is original $\bar f$ coupling as a convention).
As we have noted, we are neglecting terms from $9r_2^2 f - 6r_2 h$,
because each component of $N$ is O(100) as we will see.
The unitary matrix $U$ is the neutrino mixing matrix (up to the diagonalization matrix
of charged lepton mass matrix).

One can express $f$ matrix using a general form of $U$.
To capture the essence of the discussion, we  first use a tri-bimaximal form for $U$,
and later on add the corrections due to non-zero $\theta_{13}$.
Using the tri-bimaximal form
\begin{equation}
U = \left(
 \begin{array}{ccc}
  \frac2{\sqrt6} & \frac1{\sqrt3} & 0 \\
  -\frac1{\sqrt6} & \frac1{\sqrt3} & -\frac1{\sqrt2} \\
  -\frac1{\sqrt6} & \frac1{\sqrt3} & \frac1{\sqrt2} 
 \end{array}
\right) ,
\end{equation}
we obtain
\begin{eqnarray}
{\bf x}_1 &=& \frac{1}{\sqrt6} (c_\nu (c_1-c_2) + 2h_1, c_\nu (2c_1 +c_3) -h_2, -c_\nu(2c_2+c_3) - h_3), \\
{\bf x}_2 &=& \frac{1}{\sqrt3} (c_\nu (-c_1+c_2) + h_1, c_\nu (c_1 -c_3) +h_2, c_\nu(-c_2+c_3) + h_3), \\
{\bf x}_3 &=& \frac{1}{\sqrt2} (c_\nu (c_1+c_2) , -c_\nu c_3 +h_2, -c_\nu c_3 + h_3).
\end{eqnarray}
%
%As we have already described,
%the hierarchy of $f$ coupling is related to $\tan\theta_{23} \simeq |c_1/c_2|$. %{\bf I DONT SEE THE SIGNIFICANCE OF THIS REQUIREMENT- IT SEEMS IF WE ASSUME $C_1, C_2 << C_3$, THAT SUFFICES TO GIVE TBM AS WELL AS OUR DESIRED PATTERN FOR $f$ NEEDED IN FITTING CHARGED FERMIONS AND SUPPRESSING P-DECAY.}
%
%Actually, $\tan\theta_{23} =1$ in the tri-bimaximal case,
%and we can find that $f_{1i}$ can be small
%if
Our purpose is to find a solution where $h_1,h_2, f_{1i}$ are suppressed.
The solution can be found by assuming
\begin{eqnarray}
c_1 - c_2 = 0  &&  (n_1, n_2 \ll n_3, \ \mbox{normal hierarchy)},\\
c_1 + c_2 = 0  &&  (n_1, n_2 \gg n_3, \ \mbox{inverted hierarchy)}.
\end{eqnarray}
%for the case of normal hierarchy

In the case of normal hierarchy,
we need $v_R/n_3 \sim 10^{14}$ GeV, and therefore,
we expect $n_3 \approx 100-1000$.
Assuming $n_1 \ll n_2, n_3$, for the case of normal hierarchy, we have 
$n_2/n_3 \simeq \sqrt{\Delta m^2_{\rm sol}/\Delta m^2_{\rm atm}} \sim 0.15$.
By a choice of $c_1 = c_2$,
the $f$ matrix is obtained (assuming $h_1, h_2 \ll c_1,c_2 \ll c_3 \ll h_3$ to express in short) as
\begin{equation}
f\sim
 \frac{1}{6n_1}\left(
 \begin{array}{ccc}
  4c_\nu^2 c_1^2 \frac{6n_1}{2n_3} & -{2c_\nu^2 c_1c_3}\frac{6n_1}{2n_3} 
  & {2c_\nu c_1h_3}\frac{6n_1}{2n_3} \\
  -{2c_\nu^2 c_1c_3}\frac{6n_1}{2n_3} & {c_\nu^2 c_3^2}& {c_\nu c_3 h_3}\\
  {2c_\nu c_1 h_3}\frac{6n_1}{2n_3} & {c_\nu c_3 h_3}& {h_3^2}
 \end{array}
\right).
\end{equation}
We emphasize that the 11,12,13 elements of $f$ are suppressed by $n_1/n_3$,
which is precisely what is needed to suppress nucleon decay amplitudes.
%
%The size of $c_\nu c_3$ has to be $O(0.1)$
%as long as the muon mass comes from $f_{22}$.
%To obtain the proper value of $V_{cb}$, we need a cancellation between $f$ and $h^\prime$.
%(It can also modify the atmospheric mixing from the diagonalization of charged-lepton unless $c_e \sim +1)$.

In the case of inverted hierarchy, 
we need $|m_1|\simeq |m_2|$, and $|m_1| - |m_2| = \Delta m^2_{\rm solar} /(2m_1)$.
If $m_3 \ll m_1$, $m_1 \simeq \sqrt{\Delta m^2_{\rm atm} }$.
So, $n_1 \approx 100-1000$ in this case.
Assuming $n_3 \ll n_1,n_2$, we can obtain
%\begin{equation}
%
%
%f \sim \frac{1}{2n_3}\left(
% \begin{array}{ccc}
%  0 & 0 & 0 \\
%  0 & c_3^2 & -c_3 h_3 \\
%  0 & -c_3 h_3 & h_3^2
% \end{array}
%\right). 
%\end{equation}
%
that 11,12,13 elements of $f$ are suppressed by $n_3/n_1$
similar to the normal hierarchy case.
For example,
the 13 element of $f$ for the choice of $c_2=-c_1$ and $h_1 =0$ is
\begin{equation}
f_{13} = \frac{2c_\nu c_1 (-h_3-c_\nu c_3 +2c_\nu c_2)}{6n_1}
+ \frac{-2c_\nu c_1(h_3 +c_\nu c_1 +c_\nu c_3)}{3n_2},
%\simeq -\frac{c_1h_3}{n_1} \ \  {\rm or } \ \ \frac{c_1h_3}{3n_1}.
\end{equation}
which is equal to $-c_\nu c_1 h_3/n_1$ (for $n_1 \simeq n_2$),
$c_\nu c_1 h_3/3n_1$ (for $n_1 \simeq -n_2$), for example.

%Suppose that the $h^\prime f^{-1} h^\prime$ term is dominant in the seesaw mass
%and $ux \ll w \ll y$.
%Then, we obtain 
%\begin{equation}
%\tan\theta_{23} = \frac{c_1}{c_2},\qquad
%\tan^2\theta_{12} = \frac{c_1^2+c_2^2}{c_3^2},\qquad
%\tan\theta_{13} \sim \frac{ux}{w} \sim \frac{m_{\nu_2}}{m_{\nu_3}}.
%\end{equation}
%As a result, large solar and atmospheric mixings and small
%13 mixing (which is related to the neutrino mass hierarchy)
%can be realized in the type I seesaw.

In both cases, the $f_{1i}$ elements can be suppressed naturally,
and 
the suppression of $f_{1i}$
is related to 
the hierarchy between the VEV of ${\bf 126}$ and $v_u^2/m_\nu \sim 10^{14}$ GeV.

We note on the effect of modification from the tri-bimaximal case : ($U_{13} \neq 0$).
In the inverted hierarchy case,
the condition is just changed to
$c_2 \cos\theta_{23} + c_1 \sin\theta_{23} = 0$ (as long as $h_1 =0$).
In the normal hierarchy case,
we require $({\bf x}_1)_1=0$ to make $f_{1i} \to 0$.
In that case, $({\bf x}_3)_1 \sim c_1$ similarly to the tri-bimaximal case.
In the tri-bimaximal case, $({\bf x}_2)_1 = 0$ is satisfied (for $h_1 = 0$).
In the case of $\theta_{13} \neq 0$,
$({\bf x}_2)_1 \simeq c_1 \sin\theta_{13}/(\sin\theta_{12} \sin\theta_{23})$. %{\bf HOW DID YOU GET THIS EQUATION ?}
Therefore, the $f_{1i}$ elements are suppressed by $U_{13}^2 n_2/n_3$ and $n_1/n_3$.
Because of the approximate relation $n_2/n_3 \simeq \sqrt{\Delta m^2_{\rm sol}/\Delta m^2_{\rm atm}} \sim \theta_{13}$,
the size of the $f_{1i}$ elements is not far different from the case of $\theta_{13}=0$.

%------------------------

We solved the $f$ matrix which can reproduce the mixing angles in $U$ by Eq.(\ref{solve-f}).
As we have seen, for the solution with suppressed $f_{1i}$,
$c_1$ and $c_2$ has to be related ($|c_1| = |c_2|$ for the tri-bimaximal mixing).
If we put the mixing angles in $U$, the condition is given as $\tan\theta_{23} \simeq |c_1/c_2|$. 
Inversely speaking, if we start from a $f$ matrix with suppressed $f_{1i}$ to suppress nucleon decay naturally,
the large atmospheric mixing angle implies $|c_1/c_2| \simeq 1$. 
Interestingly, it naively implies a ``post-diction" for quark mixing : $V_{ub} \sim V_{us} m_s/m_b$.
%

%Now let us consider if the observed neutrino mixings can be naturally obtained,
%and the relation in Eq.(\ref{mass-mixing}) can be realized.
%As it can be found from the discussion above and the equation in the Appendix,
%the large atmospheric and small 13 mixing can be obtained naturally
%if $f$ matrix is hierarchical.
%Since ${\bf a}_1 = (h_1, -c_1,c_2)$, (since $c_\nu \simeq 1$) the large atmospheric mixing is related
%to $c_1 \sim c_2$ and small 13 mixing can be related to $h_1 \ll c_1,c_2$.
%
%In order to obtain a large solar mixing,
%it is necessary to satisfy
%$(U_F)_{22}^* c_\nu c_1 \sim -(U_F)_{22}^* c_\nu c_3 + (U_F)_{32}^* h_3$.
%%
%Due to the naive relation $c_\nu c_i \ll h_3$,
%there is a tension to obtain a large solar mixing 
%as long as the muon mass is generated from the 22 element of $f$ coupling
%as we have described.
%%
%A large solar mixing can arise naturally in the itemized cases described below.
%In such cases, 
%the large solar and atmospheric mixing can be obtained by $c_1 \sim c_2 \sim c_3$ 
%in the {\bf 120} Higgs coupling,
%and 13 neutrino mixing is naturally small due to the anti-symmetricity of the matrix.
%

%If the $h$ contribution can be completely insensitive in the last case, 
%the $f$ coupling in the case of inverted hierarchy can be written as
%
%\begin{equation}
%f\sim
% \frac{c_\nu^2 c_3^2}{6n_1}\left(
 %\begin{array}{ccc}
%  0 & 0 & 0 \\
%  %
%  0 & 1& 1\\
%  %
%  0 & 1& 1
% \end{array}
%\right).
%\label{typeI-f}
%\end{equation}
%

We note on the fitting of the charged fermion mass and mixing.
The size of $c_\nu c_3$ has to be $O(0.1)$
as long as the muon mass comes from $f_{22}$.
To obtain the proper value of $V_{cb}$, we need a cancellation between $f$ and $h^\prime$.
It can also modify the atmospheric mixing from the diagonalization of charged-lepton unless $c_e \sim +1$
(For example, at the SU(5)-like vacua, $c_e \sim -1$.).

Apart from the detail fit, we have showed that the suppression of $f_{1i}$ and $h_{1,2}$ is possible
to reproduce the neutrino mixings in type I seesaw.
Although the detail of the charged fermion masses and mixings may depend on the threshold corrections
(both GUT scale and weak scale) or any other possible higher order effects,
the suppression of $f_{1i}$ can generate an interesting feature of the type I seesaw.
The $f_{1i}$ elements is suppressed by $n_1/n_3$, as we have described.
In this case, the $f$ coupling can be a source of $\tau \to \mu\gamma$
without enhancing $\mu \to e \gamma$ and $\tau\to e\gamma$
if one of the SM decomposed representations in the $\bf 126$ Higgs fields is 
lighter than the unification scale \cite{Dutta:2007ai}.
The feature of this solution is obtained because the $f_{1i}$ components
can be tiny to realize the neutrino masses, 
which is suitable to suppress proton decay amplitude.
This is the main difference between type I and type II seesaw.
In the case of triplet-part-dominant type II seesaw,
it is impossible to enhance $\tau\to\mu\gamma$ around its current experimental bound 
after satisfying the bounds of 
$\mu\to e\gamma$ and $\mu$-$e$ conversion because a size of $f_{13}$ element is needed to 
generate the proper neutrino oscillation data.
In both cases, $\mu \to e\gamma$ can be generated just below the current experimental bound
via the Dirac neutrino Yukawa coupling, or left-handed Majorana neutrino couplings.
If $\tau \to \mu\gamma$ is  discovered soon, the structure in type I seesaw with suppressed
nucleon decay is preferred.

%In addition to that,
%the $f$ Majorana neutrino coupling does not generate $\mu$-$e$ flavor violation,
%and the Dirac neutrino coupling dominate the $\mu\to e\gamma$ amplitude.
%In this case, thus, the $\tau\to \mu\gamma$ can be large from the contribution
%of the $f$ Majorana neutrino coupling,
%which is the main difference between type I and type II seesaw scenario
%in the rank 1 $h$ coupling scheme to suppress proton decay amplitudes.

\section{Predictions of $\mu\to e\gamma$ decay}

In this section,
we show the predictions of the branching ratio of $\mu\to e\gamma$ decay
in the SO(10) model with proton decay suppression.

As is expressed in Eq.(\ref{off-diagonal}),
the size of the off-diagonal elements is specified by
$\kappa$,
and the numerical quantity of $\kappa$ is specified 
by the size of the coupling matrices
and the SUSY breaking parameters.
For example, if the source is the Dirac neutrino Yukawa coupling,
the numerical quantity is roughly estimated as
\begin{equation}
\kappa m_0^2 \simeq \frac{1}{8\pi^2} (Y_{\nu3}^{\rm diag})^2 (3 m_0^2 + A_0^2) \ln \frac{M_*}{M_{R3}},
\end{equation}
where 
$Y_{\nu3}^{\rm diag}$ is the 3rd eigenvalue of the Dirac Yukawa coupling matrix,
$M_*$ is a cutoff scale, $M_{R3}$ is the 3rd right-hand Majorana mass,
$m_0$ is a SUSY breaking universal scalar mass,
and $A_0$ is a universal scalar trilinear coupling.

\begin{figure}[tbp]
 \center
  \includegraphics[width=10cm]{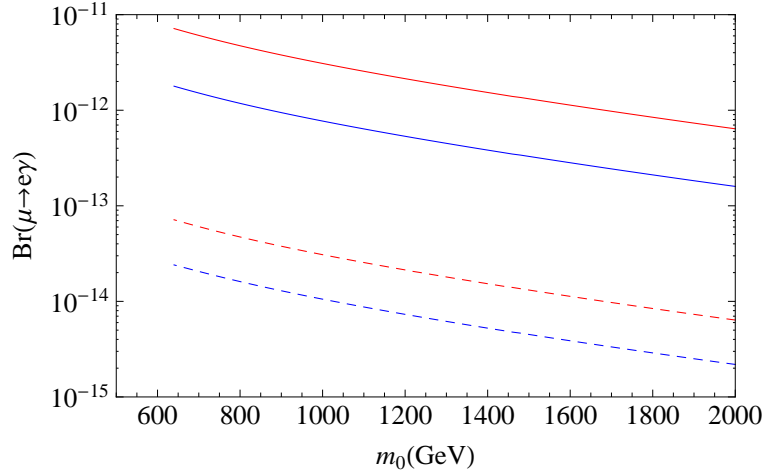}
 \caption{Branching ratio of $\mu\to e\gamma$ decay vs. the universal scalar mass $m_0$.
 We choose $M_{1/2} = 700$ GeV, and $\tan\beta=50$.
 The universal trilinear scalar coupling is chosen to satisfy $m_h = 125$ GeV.
 The current experimental bound is Br$(\mu\to e\gamma) < 2.4\times 10^{-12}$ \cite{Adam:2011ch}.
 It is expected that the decay can be observed if the branching ratio is larger than $10^{-13}$
 at the MEG experiment.
 The detail of the plots are given in the text.
}
\label{Fig2}
\end{figure}

In Fig.\ref{Fig2}, we plot the branching ratio of $\mu\to e\gamma$ decay
for various scenarios.
We choose the unified gaugino mass to be $M_{1/2} = 700$ GeV
(corresponding to the experimental lower limit of $m_{\tilde g} \simeq 1.7$ TeV for gluino mass)
to satisfy the recent LHC experimental bounds,
and $\tan\beta = 50$.
In the plot, the universal trilinear coupling $A_0$ is chosen to be
a value to make the lightest Higgs mass to be $m_h = 125$ GeV \cite{:2012gk}.
For example, for the universal scalar mass $m_0 = 1000$ GeV, we obtain $A_0 = -2000$ GeV
(In our sign convention,
 the positive value of $M_{1/2}$ takes the trilinear scalar couplings to the negative direction
 by RGEs. Namely, for a boundary condition $A_0 = 0$, 
 the scalar trilinear couplings are negative at the weak scale
 in the convention).
We note that the flavor violation is enhanced because a large magnitude of $A_0$ is needed
to obtain the Higgs mass.
For larger $m_0$, the left-right stop mixing becomes smaller (for fixed $A_0$).
Therefore, in order to obtain $m_h = 125$ GeV for a larger value $m_0$,
a larger magnitude of $A_0$ is needed, and the flavor violation is more enhanced.
As a result, the behavior of the plots are different from
 the simple dependence of the slepton masses.
Due to a large magnitude of $A_0$ and a large left-right mixing,
one of the stau masses becomes tachyonic for a small value of $m_0$.
As a result, $m_0$ has to be larger than 650 GeV for the plot.
For the given boundary conditions, squark masses are heavier than 1.5 TeV.

We choose the following four scenarios to plot:
\begin{enumerate}
\item 
(Red lines)

The flavor violating source is the neutrino Dirac Yukawa coupling,
and the unitary matrix $U$ in Eq.(\ref{off-diagonal}) is the same
as the neutrino mixing matrix.
Namely, the Dirac neutrino mass matrix can be written as
\begin{equation}
Y_\nu = U Y_\nu^{\rm diag},
\end{equation}
in the basis where the right-handed neutrino mass matrix is diagonal.
We choose $M_{R3} = 10^{13}$ GeV (red solid line)
and $3\times 10^{12}$ GeV (red dashed line).

\item
(Blue solid line)

The flavor violating source is the Majorana coupling of the left-handed 
lepton doublets and SU(2)$_L$ triplets for the type II seesaw,
and the triplet part provides the dominant contribution of the light neutrino masses
(Then, the unitary matrix $U$ is same as the neutrino mixing matrix). 
As we have explained, the coupling is the $\overline{\bf 126}$ Higgs coupling
in the SO(10) model,
and the second generation fermion masses are generated
by this coupling in the current setup for the proton decay suppression.
There is a free parameter for the $\bf 10$ and $\overline{\bf 126}$ Higgs mixing.
We choose $\bar f_3^{\rm diag} = 0.1$ for the original $\overline{\bf 126}$ Higgs coupling
without multiplying the Higgs mixing.
The numerical quantity $\kappa$ is propotional to $\bar f_3^{\rm diag}$,
and the branching ratio is roughly proportional to $\kappa^2$.
Therefore, for $\bar f_3^{\rm diag} < 0.05$, the branching ratio is smaller than
$10^{-13}$. %{\bf THE GRAPH SHOWS BR MORE THAN $10^{-13}$. DIFFERENT FROM THIS STATEMENT. EXPLAIN }
Because $f_{33}$ provides the second generation fermions mass,
and $f_{23}$ generates the quark mixing $V_{cb}$,
one can expect that the branching ratio can be more than $10^{-13}$ in this scenario.

\item
(Blue dashed line)

The flavor violating source is the Dirac neutrino Yukawa coupling, Eq.(\ref{Dirac-neutrino-Y}),
in the SO(10) models.
Contrary to the SU(5) models, the Dirac neutrino Yukawa coupling is restricted 
in the current scenarios of the SO(10) models.
There is one free parameter $c_\nu$ in the Dirac neutrino Yukawa coupling.
We choose $c_\nu = 2$ (which is the value of the SU(5)-like vacuum,
which is preferable since one obtains $c_e = -1$ and 
the Georgi-Jarskog relation can be naturally obtained).
In order to observe the $\mu\to e\gamma$ decay at the MEG experiment~\cite{Adam:2011ch}, 
one needs a larger value of $c_\nu$.
The branching ratio is roughly proportional to $c_\nu^2$.

\end{enumerate}

In addition to the above sources,
there can be a flavor violating source in the $e^c u^c H_C$ coupling,
and the colored Higgs loop can induce the off-diagonal elements
in the right-handed charged slepton mass matrix,
and can generate LFV via neutralino loop diagram.
The contribution is calculated to be less than $10^{-15}$ for the branching ratio
using the boundary condition of the SUSY particle spectrum.

In the above calculations, we assume the gaugino mass unification and the universality of the scalar masses.
However, if there is a SUSY breaking contribution from the anomaly mediation,
the gaugino mass unification can be relaxed even in the GUT modes.
In that case, the gluino mass bounds from the LHC experiments does not necessarily 
restrict the Wino and Bino masses,
and the Br($\mu\to e\gamma$) can be enhanced.

In summary, in the SU(5) GUTs with type I seesaw,
the branching ratio of the $\mu\to e\gamma$ decay can be as large as the 
current experimental bounds,
and its magnitude depends on the right-handed neutrino Majorana mass.
In the type II seesaw SO(10) model with proton decay suppression,
it is expected that the $\mu\to e\gamma$ is observed at the MEG experiments.
In the type I seesaw SO(10), on the other hand,
the branching ratio is smaller than $10^{-13}$ (for a natural size of $c_\nu$) 
and $\mu\to e\gamma$ decay may be difficult to be observed.

As is explained, the $\tau\to\mu\gamma$ decay width cannot be large enough to be observed near future
in type II seesaw SO(10) (and type I seesaw SU(5)) models,
satisfying the experimental bounds of $\mu\to e\gamma$ and $\mu$-$e$ conversion.
In type I seesaw SO(10) with proton decay suppression,
the $\tau\to\mu\gamma$ decay width can become large by the $\overline{\bf 126}$ Higgs coupling
without enhancing $\mu\to e\gamma$.
%since the coupling is given in Eq.(\ref{typeI-f}). %{\bf THIS STATEMENT IS SO INTERESTING- SHOULD WE ELABORATE A BIT ON THAT ?}
%
The $\tau\to e\gamma$ decay is not enhanced in both type I and II seesaw scenarios.
These features are important to distinguish the GUT models and the vacua of the GUT symmetry breaking.
%{\bf SHOULD WE GIVE A PLOT OR NUMBER FOR $\tau\to \mu+\gamma$}

\section{Partial lifetime of nucleon in type I and type II}

The hierarchical structure of the Yukawa coupling matrix is needed to suppress proton decay.
As we have studied, natural suppression of the proton decay amplitude is possible
in the type I seesaw
relating to the hierarchy between the GUT scale and the typical seesaw scale $\sim 10^{14}$ GeV.
In the type I seesaw, the structure is really simple
and it allows predictions for the partial decay amplitude.
which we investigate in this section.

We denote the coupling matrices in the basis where $h$ is diagonal:
\begin{equation}
h = \left(
 \begin{array}{ccc}
 h_1 && \\
 & h_2 & \\
 && h_3
 \end{array}
\right),
\quad
f =  \left(
 \begin{array}{ccc}
 u & v & x\\
 v & y & z \\
 x & z & w
 \end{array}
\right),
\quad
h^\prime =\left(
 \begin{array}{ccc}
 0 & c_1 & -c_2\\
 -c_1 & 0 & c_3 \\
 c_2 & -c_3 & 0
 \end{array}
\right).
\end{equation}
As we have explained, we assume that the $h$ coupling is rank 1 
(namely, $h_1$ and $h_2$ are irrelevant to fit fermion masses).
Then, we can choose $v = 0$ without loss of generality.
In that basis, $u\to 0$ is important to suppress proton decay amplitude.
In type II seesaw, $x$ cannot be small to fit the large solar neutrino mixing.
In type I seesaw, on the other hand,
$x$ is small and the smallness of $x$ gives a predictivity to the decay amplitudes.
In the following, we take $h_1,h_2,u,v \to 0$, but we keep $x$ to describe the difference 
between type I and II seesaw.

The left-handed proton decay amplitudes (from the $LLLL$ dimension-five operator $C_L$)
from the chargino dressing diagram
can be written as
\begin{eqnarray}
A_L(p\to K \bar \nu_\tau) &\simeq&
-\beta_p g_2^2 (x p_1 - c_2 p_5) y \cos\theta_C \sin\theta_C X_{p\to K\bar\nu}, \\
A_L(p\to K \bar \nu_\mu) &\simeq&
-\beta_p g_2^2 c_1 p_5 y \cos\theta_C \sin\theta_C X_{p\to K\bar\nu},
\end{eqnarray}
where $\beta_p$ is a hadron matrix element of proton, $\theta_C$ is the Cabibbo angle,
$p_1$ and $p_5$ are the coefficients from the colored Higgs mixing for $ff$ and $h^\prime f$ 
contributions, respectively:
\begin{eqnarray}
x_{L1} &=& X_{a4} \frac1{M_a} Y_{a5} = (M_T^{-1})_{54}, \\
x_{L5} &=& \sqrt2X_{a3} \frac1{M_a} Y_{a5} = \sqrt2(M_T^{-1})_{53} .
\end{eqnarray}
Here, $X$ and $Y$ are the diagonalization unitary matrix of the
colored Higgs mass matrix $M_T$ and
 we used the same notation in Ref.\cite{Dutta:2004zh}.
The factor $X_{p\to K\bar\nu}$ includes a loop function and chargino mixing angles $\theta_u$, $\theta_v$:
\begin{equation}
X_{p \to K\bar\nu} = (A_1-A_2) (\cos^2\theta_u H_{ue}^1 + \sin^2\theta_u H_{ue}^2)
+(A_1+A_2) (\cos\theta_u\cos\theta_v H_{ud}^1 + \sin\theta_u\sin\theta_v H_{ud}^2),
\end{equation}
\begin{equation}
A_1 =  1+ \frac{m_p}{m_{B^\prime}} (F+\frac13D),
\qquad
A_2 =  \frac{m_p}{m_{B^\prime}} \frac23D,
\end{equation}
where
$F \simeq 0.48$ and $D \simeq 0.76$ are coupling constants for interaction between
the baryons and mesons,% \cite{},
$m_p$ is the proton mass, and $m_{B^\prime}$ is an averaged baryon mass 
$m_{B^\prime} \approx m_\Sigma \approx m_\Lambda$.
The loop function is defined as
\begin{equation}
H_{ue}^\alpha = 
\frac{1}{m_{\chi_\alpha}}H\left(\frac{m^2_{\tilde u}}{m^2_{\tilde \chi_\alpha}},\frac{m^2_{\tilde e}}{m^2_{\tilde \chi_\alpha}}\right),
\end{equation}
where
\begin{equation}
H(x,y) = \frac{1}{x-y} \left(\frac{x\ln x}{x-1}- \frac{y\ln y}{y-1}   \right),
\end{equation}
$H_{ud}$ is defined similarly by replacing the slepton mass into down-type squark mass,
and
$m_{\chi_\alpha}$ is an eigenvalues of chagino masses.
We assume that SUSY breaking squark, slepton mass matrices are proportional to identity matrix
for simplicity.
We neglect the subleading contribution from $V_{cb}$ and $V_{ub}$.
The $p\to K \bar\nu_e$ is suppressed by a factor $\sim \theta_C$, but can be generated
due to the mixing between electron and muon in the basis where $h$ is diagonal.

We find
that $A_L(p \to K \bar\nu_\tau) \propto x p_1 - c_2 p_5$
and $A_L(p \to K \bar\nu_\mu) \propto c_1 p_5$.
Therefore, if $x\to 0$ (as in type I seesaw), the left-handed amplitude of $p\to K\bar\nu$ can be
suppressed by choosing a small $p_5$.
The smallness of $p_5$ is related to a vacuum selection of SO(10) breaking vacua.
In type II, on the other hand, $x$ cannot be small, and one needs to choose small values 
both $p_1$ and $p_5$ to suppress proton decay.
If there are numbers of parameter in the colored Higgs mass matrices, 
the suppression is possible. However, it is not related to the vacuum selection.

Since in the case of $x\to 0$ in type I seesaw case 
the cancellation between the Yukawa couplings for  different 
Higgs representations ($ff$ and $h^\prime f$ for example)
is not required,
the ratio of partial lifetime is predictable.
In fact, the decay amplitudes of to anti-muon are obtained as
\begin{eqnarray}
A_L (p\to \pi^0\mu^+)
&\simeq&
\beta_p g_2^2 c_1 p_5 y \sin\theta_C \sin\theta_{uc} A_3 X_{p\to \mu +X}, \\
A_L (p\to K^0\mu^+)
&\simeq&
-\beta_p g_2^2 c_1 p_5 y \cos\theta_C \sin\theta_{uc} A_4 X_{p\to \mu +X},
\end{eqnarray}
where
\begin{equation}
X_{p\to\mu+X} = \cos\theta_{u} \cos\theta_{v} (H_{d\nu}^1+ H_{ue}^1) + 
\sin\theta_{u} \sin\theta_{v} (H_{d\nu}^2+H_{ue}^2) ,
\end{equation}
and
\begin{equation}
A_3 =  \frac{1}{\sqrt2} (1+F+D), \qquad A_4 =  1+ \frac{m_p}{m_{B^\prime}} (F-D).
\end{equation}
The mixing angle $\theta_{uc}$ is an angle between $u$ and $c$ quarks in the basis
and therefore, $\theta_{uc} \sim \sqrt{m_u/m_c}$.
Therefore,
we obtain the ratio of partial decay width in type I seesaw as
\begin{equation}
\frac{\Gamma (p\to K\mu)}{\Gamma (p\to K \bar\nu)}
\simeq
\frac{|c_1|^2}{|c_1|^2+|c_2|^2}
\frac{\sin^2\theta_{uc}}{\sin^2\theta_C}
\left|\frac{A_4 X_{p\to \mu+X}}{X_{p\to K\bar\nu}}\right|^2.
\end{equation}
The maximal atmospheric mixing in the type I seesaw requires
$|c_1| \simeq |c_2|$, and the ratio is predictive (up to the sfermion mass spectrum).
In the case of type II seesaw, there is an additional parameter $(x- c_2 p_5/p_1)$,
and the ratio cannot be predicted.
We comment that 
\begin{equation}
\frac{\Gamma(p\to \pi^0 \mu^+)}{\Gamma(p\to K^0\mu^+)}
\simeq \sin^2\theta_C \frac{A_3^2}{A_4^2}
\end{equation}
is obtained for both type I and II
since only $h^\prime f$ contribution is dominated.
For the decay to the third generation lepton 
(because tau lepton is heavier than proton, the decay to the $\nu_\tau$ is the only case),
$ff$ piece also contributes if $f_{13} = x$ is not suppressed.
Therefore,
the ratio of partial decay width to neutrino and anti-lepton is not predictable in the type II,
while it is predictable for type I seesaw.
The same situation occurs for $n \to \pi \bar\nu$ decay.

In the above expressions,
we neglect the contribution from the 
right-handed dimension-five proton decay operator $C_R$.
They can contribute to the decays into $\bar\nu_\tau$,
and can alter the prediction of the ratio of partial decay widths.
The $C_R$ contribution (for $h_1,h_2,u,v,x\to 0$) can be obtained as
 \begin{eqnarray}
A_R(p\to K \bar\nu_\tau)  
&=& \alpha_p y_c y_\tau (A_1 c_1 c_2  x_{R8} \cos\theta_C
+ A_2 (c_1 c_3 p_{R8} - c_2 y p_{R7} + c_1 z p_{R5}) \sin\theta_C) \nonumber \\
&&\times(H_{ue}^2 \cos\theta_u \cos\theta_v + H_{ue}^1 \sin\theta_u \sin\theta_v),
\end{eqnarray}
where $\alpha_p$ is a hadron matrix element, and 
\begin{eqnarray}
p_{R5} &=& -\sqrt2 X_{a2} \frac1{M_a} (Y_{a5}-\sqrt2 Y_{a6}) = -\sqrt2 (M_T^{-1})_{52} + 2 (M_T^{-1})_{62}, \\
p_{R7} &=& -\sqrt2 X_{a4} \frac1{M_a} (Y_{a3}-Y_{a2}) = -\sqrt2 ((M_T^{-1})_{34} - (M_T^{-1})_{24}), \\
p_{R8} &=& 2 X_{a2} \frac1{M_a} (Y_{a3}-Y_{a2}) = (M_T^{-1})_{32} - (M_T^{-1})_{22}. 
\end{eqnarray}
For a large $\tan\beta \sim 50$,
since $c_i$ and $y,z$ have to be large to fit the down-type quark Yukawa couplings,
$C_R$ contribution can make the decay width comparable to the 
current experimental bounds,
and the simple relation shown previously can be disturbed by the second term ($A_2$ contribution).
If the Higgsino mass is much heavier than wino mass, the $C_R$ contribution
is suppressed rather than $C_L$ contribution.
We note that if we do not assume $h_1,h_2,u,v\to 0$,
the magnitudes of $C_R$ contribution can excess the experimental bounds even for $\tan\beta \sim 5$
and we should not neglect the $C_R$ contribution.

\section{Conclusion}

We study the Yukawa texture in a renormalizable SUSY SO(10) GUT model for  neutrino masses that gives a ``large'' $\theta_{13}$ while at the time suppressing nucleon decay without invoking cancellation between coupling parameters. We consider cases with both type I and type II seesaw separately. 
In the type II seesaw scenario, we find that $\theta_{13}>0.05$ radians irrespective of the detail of fitting of fermion masses and mixings (i.e. independent of the number of parameters) and that the measured value of 13 mixing angle prefers $\delta \sim \pi$
for the CP violating phase in the neutrino oscillations. We then study the predictions of these scenarios for lepton flavor violation.
The branching ratio of $\mu\to e\gamma$ can be as large as the current experimental bound,
and should be observed very soon.
The Br($\tau\to\mu\gamma$) is predictable, and  is about $O(10^{-10})$ in the type I seesaw case,
if the bounds of $\mu\to e\gamma$ and $\mu$-$e$ conversion are satisfied.

In type I seesaw scenario, observed $\theta_{13}$  can be accommodated naturally.
We find that 1-2 and 1-3 flavor violation from the FCNC source is suppressed compared 
to the type II scenario
if the proton decay is suppressed naturally.
This gives
a larger value of Br($\tau\to\mu\gamma$) 
%without conflicting with the bound of $\mu\to e\gamma$ and $\mu$-$e$ conversion,
while $\mu\to e\gamma$ and $\tau\to e\gamma$ are suppressed.
If the $\tau\to\mu\gamma$ decay is observed soon then definitely it will point towards a  type I scenario in this kind of SO(10) model and 
the prediction for Br($\mu\rightarrow e\gamma$) can be checked for confirmation. Detailed numerical study of the model for fermion mass fits as well as lepton flavor violation is currently under way.
The ratio of the partial decay width of proton to $K \mu$  and $K\bar\nu$ is predicted 
(up to uncertainties from sfermion masses) in this scenario.

%The flavor structure can be probed by precise measurements of neutrino oscillation parameters
%(with CP phase) and the lepton flavor violations.

The next generation of the baryon number violating nucleon decay experiments
at HyperKamiokande along with the information about the low energy SUSY states from the LHC may be used to distinguish between  the different scenarios.
%Before the HyperKamiokande starts, the LHC experiments will judge the low energy SUSY,
%which is also an important building blocks to draw the whole picture.

\section*{Acknowledgments}

%\noindent

The work of  B. D.
is supported in part by the DOE grant DE-FG02-95ER40917. The work of Y.M. is supported by the Excellent Research Projects of
 National Taiwan University under grant number NTU-98R0526. The work of R.~N.~M.  is supported by the US National
Science Foundation under grant No. PHY-0968854.

\end{document}